\documentclass[superscriptaddress, preprint, amsmath, amssymb, aps, physrev]{revtex4-2}

\usepackage{float}
\usepackage[english]{babel} 
\usepackage{subcaption} 
\DeclareUnicodeCharacter{2212}{-}
\usepackage{booktabs}  
\usepackage{siunitx}   
\usepackage{amsmath}
\usepackage{chemformula}
\usepackage{mhchem}
\usepackage[utf8]{inputenc}

\usepackage{chemformula}

\begin{document}

\title{Tailoring the Electronic Properties of Monoclinic (In$_x$Al$_{1-x}$)$_2$O$_3$ Alloys via Substitutional Donors and Acceptors
}


\author{Mohamed Abdelilah Fadla}
\affiliation{School of Mathematics and Physics, Queen's University Belfast, University Road, Belfast BT7 1NN, UK}
\email{m.fadla@qub.ac.uk}
\author{Myrta Gr\"{u}ning}
\affiliation{School of Mathematics and Physics, Queen's University Belfast, University Road, Belfast BT7 1NN, UK}
\affiliation{European Theoretical Spectroscopy Facility}
\author{Lorenzo Stella}
\affiliation{School of Mathematics and Physics, Queen's University Belfast, University Road, Belfast BT7 1NN, UK}

\begin{abstract}

Ultra-wide bandgap semiconductors such as $\beta$-$\mathrm{Ga_2O_3}$ are ideal materials for next-generation power electronic devices. Electronic and mechanical properties of $\beta$-$\mathrm{Ga_2O_3}$ can be tuned by alloying with other sesquioxides, notably $\mathrm{Al_2O_3}$ and $\mathrm{In_2O_3}$. Moreover, by tuning the In content of a (In$_x$Al$_{1-x}$)$_2$O$_3$ alloy, its lattice constants can be matched to those of Ga$_2$O$_3$, while preserving a large conduction-band offset.  In view of potential applications to $\beta$-$\mathrm{Ga_2O_3}$-based heterostructure, we performed atomistic modelling of (In$_x$Al$_{1-x}$)$_2$O$_3$ alloys using density functional theory to investigate thermodynamic and electrical properties of conventional group IV dopants (Si, Sn, C, Ge), alternative metal donors (Ta, Zr, Hf), and acceptors (Mg, Zn, Cu).
The hybrid Heyd–Scuseria–Ernzerhof functional (HSE06) is used to accurately quantify the defect formation energies, ionization levels, and concentrations over a wide range of experimentally relevant conditions for the oxygen chemical potential and temperature. In our atomistic models, Hf and Zr show favourable properties as alternative donors to Si and other group IV impurities, especially under oxygen-poor conditions. Our findings also suggest that acceptors Mg, Zn, and Cu, while they cannot promote p-doping, can be still beneficial for the compensation of unintentionally n-doped materials, \emph{e.g.}, to generate semi-insulating layers and improve rectification.
  
\end{abstract}
\maketitle
\section{Introduction}

Ultra-wide bandgap (UWBG) semiconductors like $\mathrm{SiC}$, $\mathrm{GaN}$, $\beta$-$\mathrm{Ga_2O_3}$ and diamond have been extensively studied due to their attractive properties, including high breakdown electric field, thermal stability, and radiation hardness.\cite{Tsao2018} 
Among the UWBG materials, $\beta$-$\mathrm{Ga_2O_3}$, with a bandgap of 4.7 eV and breakdown electric field of about 8 MV/cm, has emerged as a promising material for next-generation power electronic devices. Additionally, high-quality single crystals of  $\beta$-$\mathrm{Ga_2O_3}$ can be produced from the melt, as for silicon, facilitating the manufacturing of $\beta$-$\mathrm{Ga_2O_3}$ wafers.\cite{galazka_bulk_2014,tadjer2022toward} 

Performance and reliability of UWBG in electronic devices critically depend on identifying and controlling dopants and defects, as it is essential for tailoring their electrical properties. \cite{dreyer2024defects,lany_defect_2018,mccluskey_point_2020,tadjer_editors_2019} 
Such properties have been thoroughly investigated for $\beta$-$\mathrm{Ga_2O_3}$ \cite{varley_oxygen_2010,peelaers_deep_2019,McCluskey2020,zhang2020recent,kaewmeechai_structure_nodate,frodason_diffusion_2023} and a current limitation is the absence of viable positive (hole) doping strategies, which precludes the development of bipolar junctions and rectifiers. A recent study on n-type doping of $\mathrm{(In_{x}Al_{1-x})_{2}O_{3}}$ has identified Si as an effective dopant, \emph{i.e.}, a shallow defect with low formation energy, while Sn and Ge become competitive at larger In fractions.\cite{seacat2024achieving}  This conclusion also holds for  $\beta$-$\mathrm{Ga_2O_3}$, although previous studies have focused predominantly on the formation energies and ionization levels of group-IV elements, alternative dopants and acceptors can be considered. For instance, the possibility of doping with transition metals such as Zr, Ta and Hf, as well as potential acceptors like Mg, Zn, and Cu, has garnered limited interest for these $\mathrm{(Al_{x}In_{1-x})_{2}O_{3}}$ alloys compared to $\beta$-$\mathrm{Ga_2O_3}$.\cite{hommedal_broad_2024,gustafson_zn_2021,kananen_electron_2017,kyrtsos_feasibility_2018,karbasizadeh_transition_2024}  Indeed, alloying  $\beta$-$\mathrm{Ga_2O_3}$ with a gate oxide is commonly used to engineer the band offset and reduce the lattice mismatch and crack formation at their interfaces.\cite{fadla2024effective, mu_first-principles_2020}  However, the investigation of the defect energies and dopability of alloys remain relatively unexplored.  It is still unclear whether these impurities remain effective across varying indium and aluminum concentrations.

In this work, we aim to systematically investigate substitutional defects in $\mathrm{(In_{x}Al_{1-x})_{2}O_{3}}$, to determine its potential for n-type dopability and explore potential acceptor dopants. Indeed, substitutional acceptors can be used to overcome the tendency of n-type doping to suppress the surface electron accumulation layer (SEAL) and enhance rectification properties.\cite{splith_numerical_2021} In particular, the following candidate donors, Si, Sn, C, Ge (group IV), as well as the promising alternative donors, including  Ta, Zr, and Hf (transition metals), along with the potential acceptors, Mg, Zn, Cu. By employing the Heyd–Scuseria–Ernzerhof hybrid functional (HSE06), our objectives are to systematically determine the defect formation and ionisation energies. In this way, we are able to classify the defect behaviour, investigate their compensation mechanism and critical composition.  Our findings can inform the characterisation and engineering of emerging materials towards more energy-efficient and reliable electronic devices.

\section{Methodology}

In this study, density functional theory (DFT) calculations were employed with the projector augmented wave (PAW) method as implemented in the Vienna \textit{ab initio} simulation package (VASP).\cite{kresse_efficient_1996,kresse1996efficiency,kresse1993ab, blochl_projector_1994}  All calculations were performed (unless otherwise specified) using the Heyd–Scuseria–Ernzerhof hybrid functional (HSE06) with 32\% Hartree–Fock exact exchange.\cite{10.1063/1.1564060,krukau_influence_2006} A plane wave energy cutoff of 600 eV was utilised, with an energy tolerance set at $10^{-6}$ eV. Ground-state geometries were fully optimised using a convergence criterion of 0.01 eV/Å for the largest force components. 

For bulk primitive cells, to minimize Pulay stress,  a high cutoff energy was used and sequential optimisation steps were performed, refining the final structure of the previous step, until the forces converged within a single ionic step. Tetrahedron smearing \cite{blochl1994improved} was employed to improve the convergence of the density of states (DOS).

All calculations were performed in a 120-atom supercell produced from $1 \times 3 \times 2$  expansion of C2/m conventional unit cell using a $\Gamma$-centered $2 \times 2 \times 2$ k-point mesh. Full cell optimisation was performed for the pristine structures $\mathrm{(In_{x}Al_{1-x})_{2}O_{3}}$ with $x=0$, $0.5$, and $1$, see Table S2 for the corresponding lattice parameters. For each defective structure and dopant charge state, all atomic positions were relaxed, while fixing the cell shape and volume. The same convergence criterion was used for both the pristine and defective structures, \emph{i.e.}, 0.01 eV/Å for the largest force component. Spin polarization was taken into account.
The defect formation energies, $E_{\mathrm{f}}\left[X^{q}\right]$, are computed using the standard approach,\cite{van_de_walle_first-principles_2004,freysoldt2014first} as

\begin{equation}
E_{\mathrm{f}}\left[X^{q}\right]=E_{\text {tot }}\left[X^{q}\right]-E_{\text {tot }}[\text { bulk }]-\sum_i n_{i} \mu_{i}+q (E_{\mathrm{F}} + E_{\text{VBM}}) + E_{\text {corr }}^q,
\label{eq:1}
\end{equation}

where $E_{\text{tot}}[X^q]$ is the total energy of the charged, defective supercell, and  $q$  is the number of electrons either added to ($q < 0$) or removed from ($q > 0$) the supercell.$E_{\text{tot}}[\text{bulk}]$ is the total energy of a pristine supercell (the same size as the defective one) and $\mu_i$ is the chemical potential of the substituted i$^{th}$ species. For each species,  $n_i$ gives the number of atoms either added (if $n_i > 0 $),  or removed, if ($n_i < 0$) in the defective supercell. $E_{\text{VBM}}$ is the valence band maximum obtained from the band structure of the pristine bulk material and $E_{\mathrm{F}}$ is the Fermi level, measured from the VBM.  The last term, $E_{\text {corr}}$, corrects for image-charge interactions, which is needed to account for spurious finite-size supercell effect. For this purpose, the Kumagai-Oba (eFNV) charge correction scheme was employed.\cite{kumagai_electrostatics-based_2014,freysoldt_fully_2009}

The chemical potentials of removed or added atoms depend on experimental conditions. By varying the chemical potential, different experimental scenarios can be explored. However, the chemical potential is subject to specific bounds to prevent the formation of secondary phases. These limits were determined based on the thermodynamic equilibrium condition, considering all relevant secondary phases(see Table S1 in the Supporting Information). Initial structures of the secondary phases were obtained from the Materials Project, \cite{Jain2013} and then relaxed using the same computational parameters as those applied to pristine structures.

The dominant entropic contribution is configurational, \textit{i.e.}, from the number of equivalent ways in which a defect can be set in a given supercell. Configurational entropy is crucial in determining the equilibrium defect concentration and its temperature dependence. Defect concentration is calculated by minimising the Gibbs free energy at a given temperature and pressure, with respect to  the number of defects. The equilibrium concentration --- in the dilute limit, \textit{i.e.}, neglecting defect interactions --- is given by:

\begin{equation}
\label{eq:2}
C_{X^{q}} = g N_L \exp\left(-\frac{E_{\mathrm{f}}\left[X^{q}\right]}{k_B T}\right)
\end{equation}

where $N_L$ is the number of lattice sites, $g$ is the degeneracy factor which accounts for structural and spin degeneracies, $k_B$ is the Boltzmann constant, and $T$ is the lattice temperature. In this framework, the entropic contribution was simply included by considering the multiplicity of the defect lattice site and their associated degeneracy.\cite{mosquera-lois_imperfections_2023}

The \texttt{doped}\cite{Kavanagh2024} defect simulation package was used for defect structure generation, parsing the calculations, and analyzing the results. The \texttt{ShakeNBreak}\cite{Mosquera-Lois2022} package was employed to identify energy-lowering distortions using local bond distortions and search for ground and metastable defect structures, thus avoiding incorrect predictions of formation energy and ionisation levels. The phonon frequencies were computed using the finite difference method as implemented in \texttt{Phonopy}. \cite{phonopy-phono3py-JPCM, phonopy-phono3py-JPSJ}

\newpage
\section{Results}
\begin{figure}[H]
     \centering
\includegraphics[width=12cm]{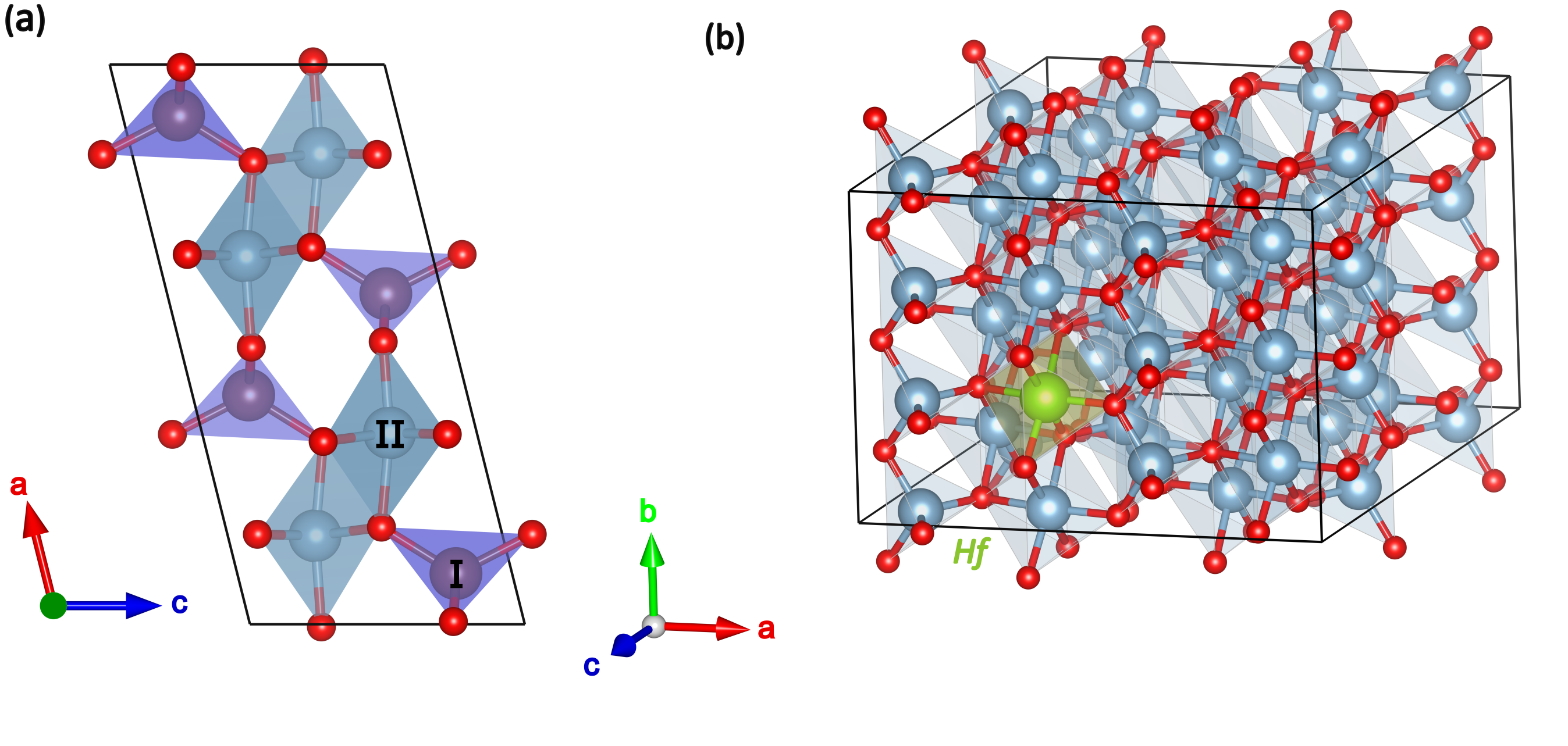} 
     \caption{(a) Conventional monoclinic unit cell showing non-equivalent tetrahedral (I) and octahedral (II) configurations. (b) Supercell representation of a  $1 \times 3 \times 2$ expanded conventional unit cell containing one Hf substitutional impurity at an octahedral site (II).}
     \label{fig-0}
\end{figure}

The monoclinic phase of  $\mathrm{Al_2O_3}$ has two non-equivalent cation sites, with tetrahedral and octahedral coordination, respectively, as shown in Fig.~\ref{fig-0}(a). Fig.~
\ref{fig-0}(b) shows the pseudo-cubic $1 \times 3 \times 2$ supercell representation used in the defect modelling. In atoms prefer to occupy octahedral sites, while Al atoms prefer to occupy tetrahedral sites. At x = 0.5, all Al atoms preferentially occupy tetrahedral sites and all In atoms preferentially occupy octahedral sites. Hence, the lowest energy configuration is an ordered alloy. 
Dynamic stability of the ordered structures, $(\mathrm{In_{x}Al_{1-x}})_2\mathrm{O_3}$ where $x=0$, $0.5$, and $1$, is assessed from the phonon dispersion, as shown in Fig. S2. No imaginary frequency modes were detected.  Band gaps, experimental (from Ref. \cite{aahman1996reinvestigation})  and optimised lattice constants are also given in Table S2.
We performed an initial screening of potential substitutional sites to exclude high-energy configurations. Detailed analyses of these preferential sites and their implications are discussed below in the relevant sections.

\subsection{Group IV substitutional dopants}

\begin{figure}[H]
     \centering
\includegraphics[width=17cm]{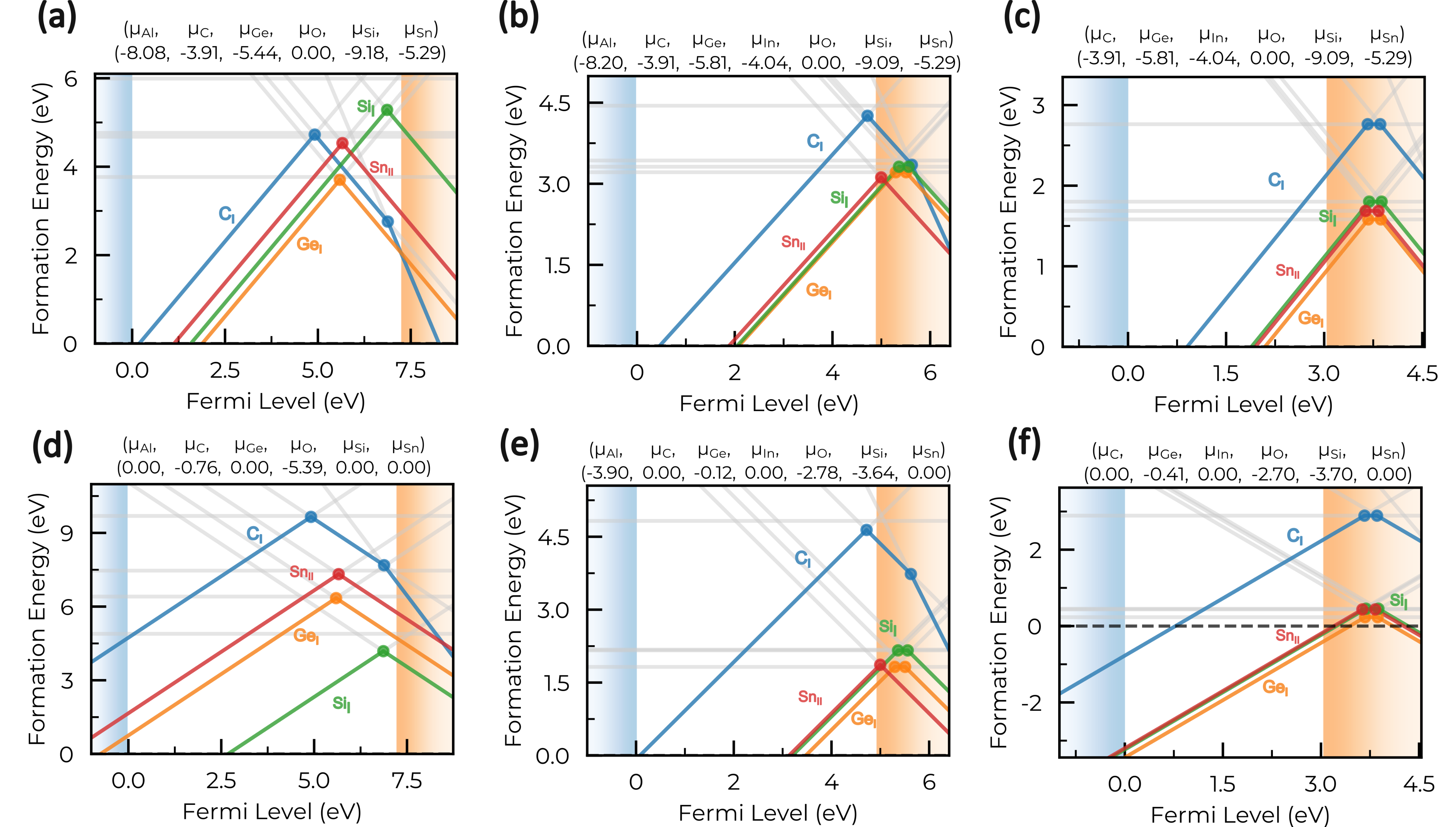} 
     \caption{Formation energies of substitutional defects C, Ge, Si, and Sn (in blue, orange, green, and red, respectively) in Al$_2$O$_3$ (a, d), \ce{(In_{0.5}Al_{0.5})_2O3} (b, e), and In$_2$O$_3$ (c, f) as a function of the Fermi level. Panels (a, c, d) correspond to oxygen-rich conditions, while panels (b, e, f) correspond to oxygen-poor conditions. The faded lines indicate regions where the defect charge states are unstable. Chemical potentials are reported at the top of the panels. For the sake of simplicity, the values are given with respect to their standard state, \emph{e.g.}, $\mu_O=0.00$ means extreme oxygen rich condition, see SI.}
     \label{fig-1}
\end{figure}

The total energies for Group-IV substitutional dopants are computed at both octahedral and tetrahedral sites. The defect formation energies as a function of the Fermi level for the lowest-energy configurations --- Si, C, and Ge in tetrahedral sites, and Sn in octahedral sites --- are shown in Fig.~\ref{fig-1} under both oxygen-poor and oxygen-rich conditions.
The slope of the coloured lines gives the charge state of each defect: a negative slope corresponds to a negatively charged state ($-$), a positive slope to a positively charged state ($+$), and a flat slope to a neutral state (0). The formation energy values are computed relative to reference states of each element, considering different competing phases as indicated by the chemical potentials given above the panels.

For Al$_2$O$_3$  (\textit{i.e.}, $x=0$) under oxygen-poor conditions, Si substitution (green line) generally displays the lowest formation energy across a broad range of Fermi levels. For In higher concentrations (\textit{i.e.}, (In$_x$Al$_{1-x}$)$_2$O$_3$ where $x=0.5$ and $x=1$), Si, Ge, and Sn dopants have comparable formation energies, but their stability varies depending on the Fermi level. Carbon dopant, on the other hand, has relatively high formation energies. (In$_{0.5}$Al$_{0.5}$)$_2$O$_3$ is a mixed-cation oxide, so the presence of both Al and In sites adds complexity to defect formation, but a trend similar to In$_2$O$_3$ is observed. Under oxygen-poor conditions, substitutional dopants like Sn$_{\text{In}}$ (and to a lesser extent Si$_{\text{In}}$, Ge$_{\text{In}}$) have lower formation energies at high In composition --- even negative energies under extreme In-rich conditions ---. Under oxygen-rich conditions, the formation energies of these dopants can be slightly larger.

\begin{figure}[H]
     \centering
\includegraphics[width=14cm]{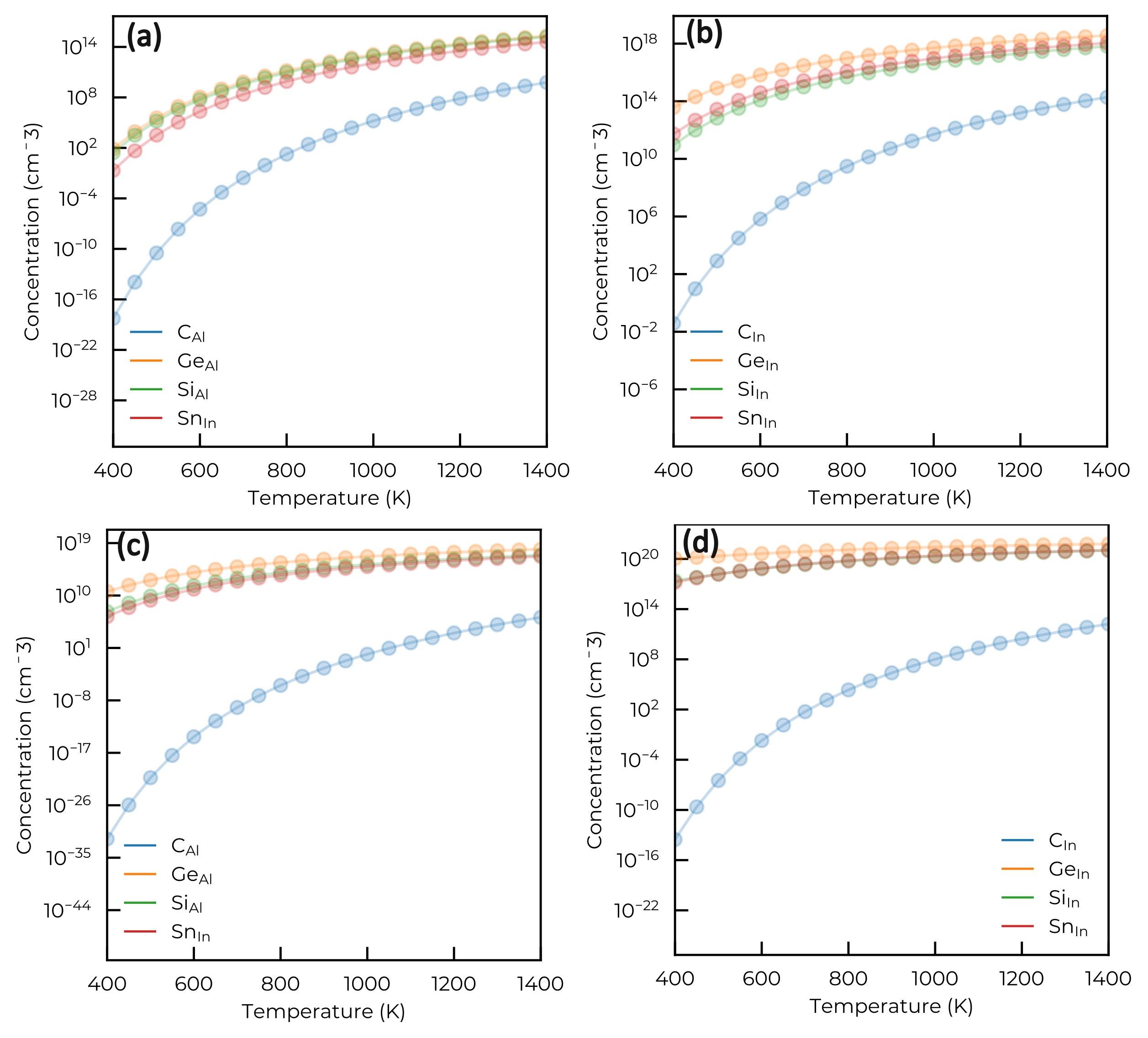} 
     \caption{Temperature sensitivity of the defect concentration of substitutional dopants C, Si, Ge, and Sn in (a, c) \((\text{In}_x\text{Al}_{1-x})_2\text{O}_3\) for \(x = 0.5\), and (b, d) \(x = 1\) over an indicative range of temperature. Panels (a and b) correspond to oxygen-rich conditions, while (c and d) correspond to oxygen-poor conditions.}
     \label{fig-2}
\end{figure}

Next, the concentration of different group IV substitutional defects (C\textsubscript{In}, Ge\textsubscript{In}, Si\textsubscript{In}, Sn\textsubscript{In}) for both \ce{In2O3} and \ce{(In_{0.5}Al_{0.5})2O3} is illustrated in Fig.~\ref{fig-2}. 
The defect concentrations for Al$_2$O$_3$ are excluded, as they consistently have low values across all dopants, consistent with the high formation energy presented in Fig.~\ref{fig-1}. Ge and Sn substitutions tend to have the highest concentrations (\(\sim10^{19} \text{ cm}^{-3}\) at 600\textdegree C), which remain nearly constant under oxygen-poor conditions, across the temperature range, compared to C, which exhibits much lower concentrations, with a significant increase as temperature rises. Defect concentrations are presented as a function of O chemical potential with an annealing temperature of [600-1000]\textdegree C, as shown in Fig.S3. In$_2$O$_3$ exhibits significantly higher dopant concentrations for all doping elements considered in this work. Alloying with Al yields a reduction in defect concentration, as shown in the case of the ordered alloy, \ce{(In_{0.5}Al_{0.5})_2O3}. With respect to this alloy, the defect concentration decreases dramatically in higher Al compositions.

\begin{figure}[H]
     \centering
\includegraphics[width=10cm]{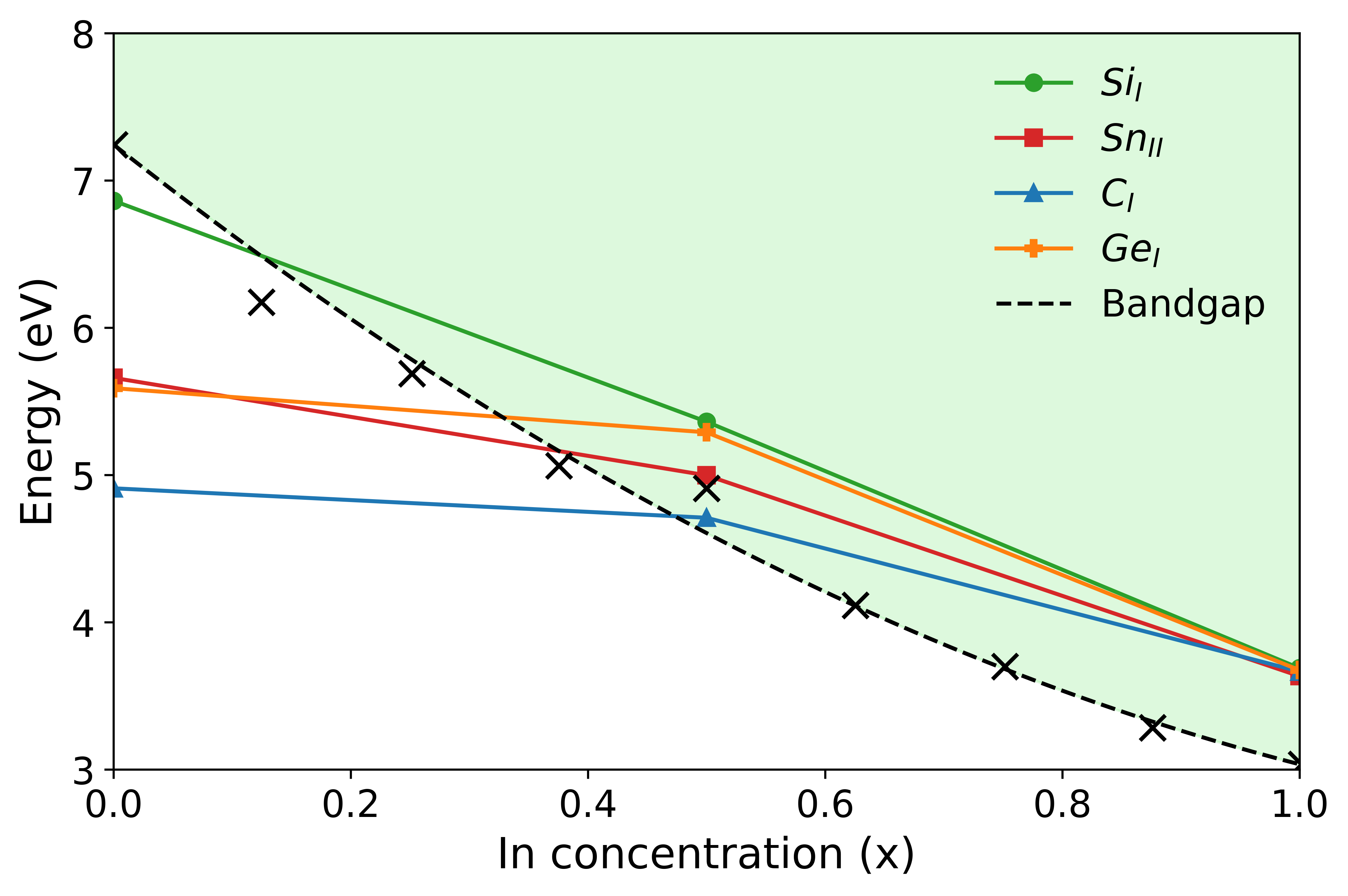} 
     \caption{First charge-state transition levels (+/0) or (+/−) for the C, Ge, Si, and Sn dopants in \ce{(In$_x$Al$_{1-x}$)$_2$O$_3$} alloy. Explicit calculations for $x=0$, $0.5$, and $1$ are indicated with points, which are connected by linear interpolation. Black crosses represent the calculated indirect bandgap values, while the dashed black line shows a quadratic fit of the bandgap. The subscripts `I' and `II' denote the tetrahedral and octahedral positions, respectively.}
     \label{fig-4}
\end{figure}

The $(+/0)$ and DX levels are interpolated between the directly calculated points at $x = 0$, $x = 0.5$, and $x = 1$ and compared with the CBM, as illustrated in Fig.~\ref{fig-4}. For the dopants C, Ge, Si, and Sn in tetrahedral (I) and octahedral (II) positions across the \ce{(In$_x$Al$_{1-x}$)$_2$O$_3$} alloy composition. 

The intersection of each dopant's transition level with the CBM gives the compositional threshold (or critical composition) beyond which the dopant introduces deep levels within the bandgap, precluding an efficient n-type doping. Specifically, (\( \mathrm{Si}_I \)) has a critical composition of $x = 0.124$, corresponding to a bandgap of 6.490~eV, indicating a transition to a deep defect at a notably low In concentration. (\( \mathrm{Ge}_I \)) exhibits a critical composition at $x = 0.328$ (bandgap = 5.393~eV), followed by (\( \mathrm{Sn}_{II} \)) at $x = 0.376$ (bandgap = 5.163~eV), while (\( \mathrm{C}_I \)) remains a shallow donor up to $x = 0.473$ (bandgap = 4.720~eV) before the onset of a deep defect state. 

\begin{figure}[H]
     \centering
\includegraphics[width=16cm]{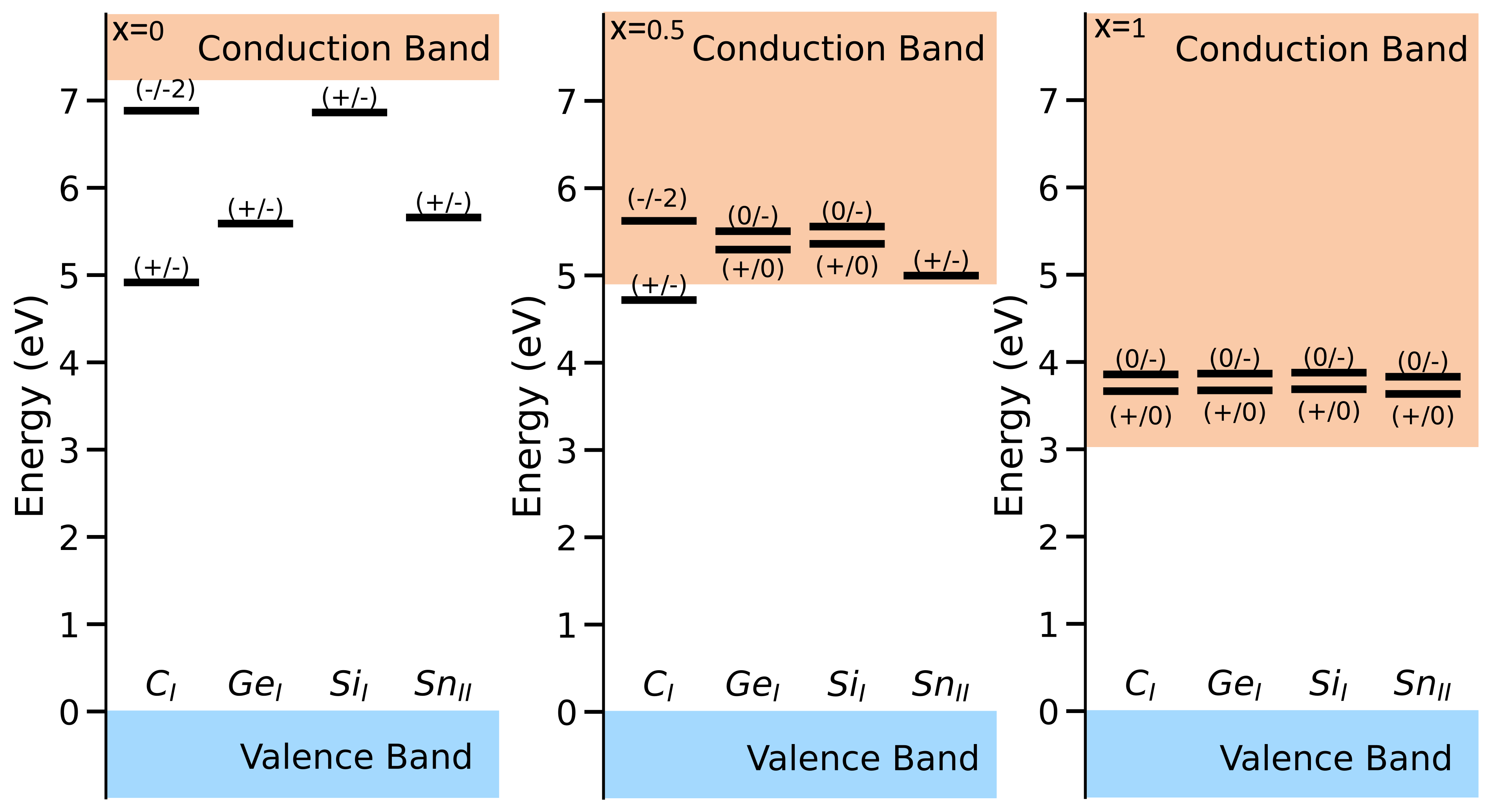} 
     \caption{The calculated charge-state transition levels for group IV substitutional dopants in \ce{(In$_x$Al$_{1-x}$)$_2$O$_3$} alloys, for   $x=0$ (left), $x=0.5$ (center), and $x=1$ (right). Numerical values are provided in Supplementary Table~S3.}
     \label{figure-1}
\end{figure}

Group IV dopants for both In$_2$O$_3$ and \ce{(In_{0.5}Al_{0.5})_2O3} are stable in the positive charge state for the entire Fermi level, all dopant transition levels occur above the conduction band minimum (CBM),  except C (for \ce{(In_{0.5}Al_{0.5})_2O3}) which exhibits +/- transition states slightly below (within \(0.2\)  eV) the band gap. In Al$_2$O$_3$, all dopants have charge-state transitions that occur within the band gap. 
The charge-state transition level values for group IV dopants are illustrated in Fig~\ref{figure-1}.

\subsection{Transition metals substitutional dopants}

\begin{figure}[H]
     \centering
\includegraphics[width=17cm]{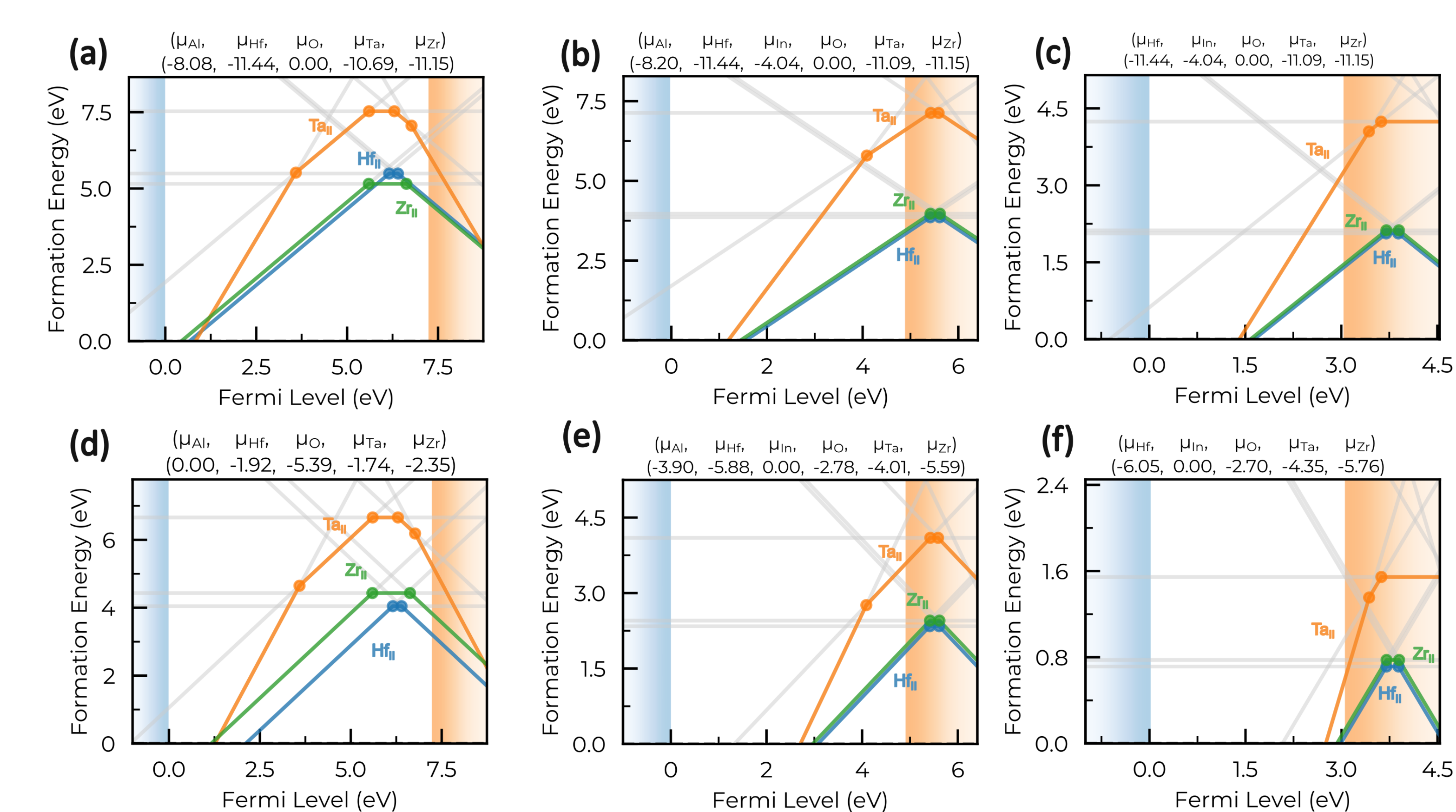} 
     \caption{As Figure~\ref{fig-1}, but for metal defects, Hf, Ta, and Zr (in blue, orange, and green, respectively).}
     \label{fig-5}
\end{figure}

In all cases considered in this work, Ta, Hf, and Zr occupy octahedral sites in their lowest-energy defect configurations. The corresponding formation energies as a function of the Fermi level are shown in Fig.~\ref{fig-5} under both oxygen-poor and oxygen-rich conditions. The dopants Hf, Ta, and Zr are represented by different coloured lines (blue, orange, and green, respectively).

Hf dopant generally displays the lowest formation energy across a broad range of Fermi levels, although Zr dopant shows comparable or slightly lower formation energies for Al$_2$O$_3$ under oxygen-rich conditions,  depending on the Fermi level. 
 
At lower Al composition, in both\ce{(In_{0.5}Al_{0.5})_2O3} and \ce{In2O3} under oxygen-poor conditions, substitutional dopants Hf$_{\text{In}}$ and Zr$_{\text{In}}$ display favourably low formation energies. For In$_2$O$_3$ under extreme In-rich conditions,  these dopants maintain low, nearly negative formation energy  for all Fermi levels. Similar to group IV, the formation energies for these dopants under oxygen-rich conditions are slightly higher.

\begin{figure}[H]
     \centering
\includegraphics[width=13cm]{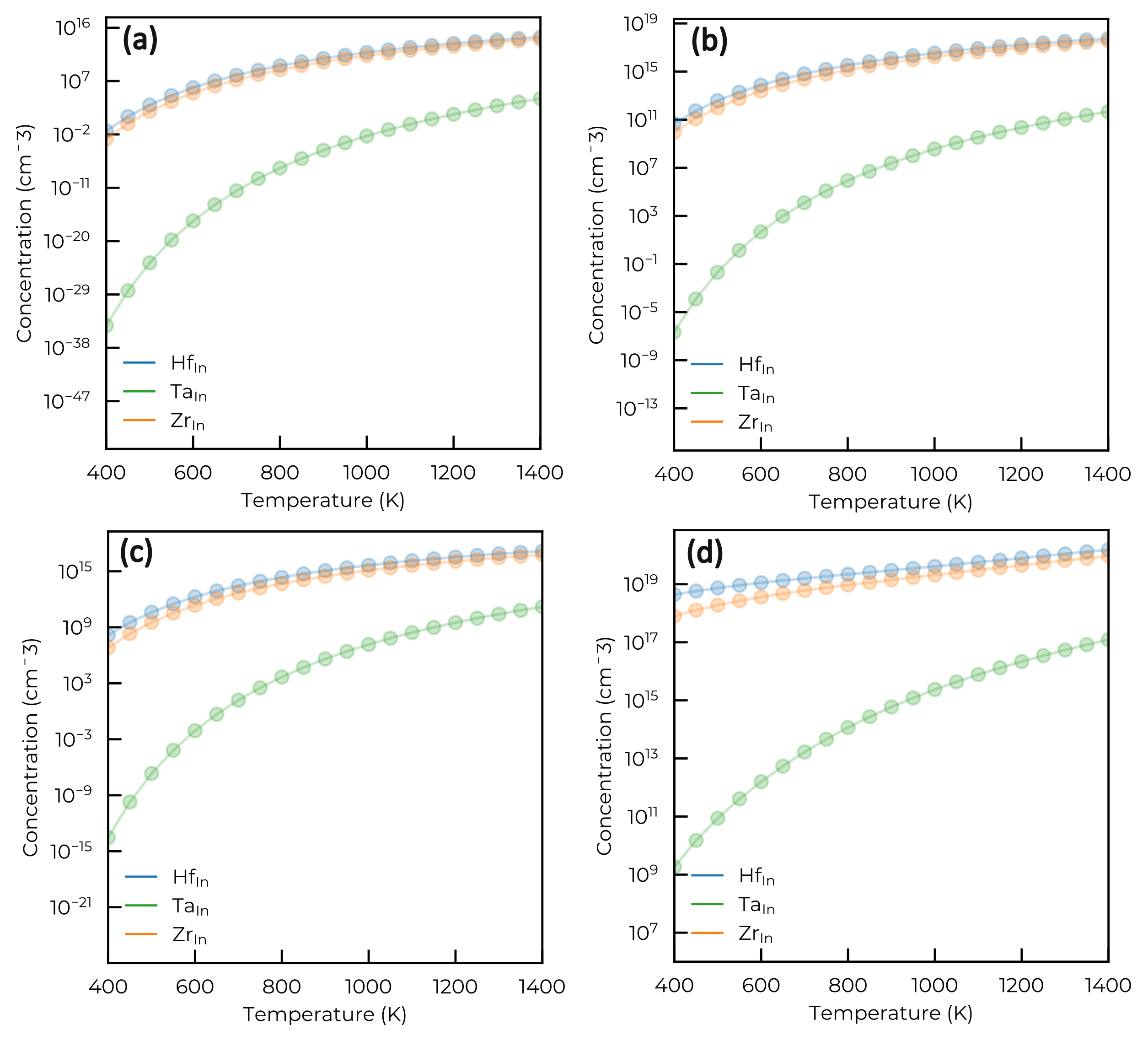} 
     \caption{Temperature sensitivity of the defect concentrations in \ce{(In_{0.5}Al_{0.5})_2O3} (a, c) and \ce{In2O3} (b, d) over an indicative range of temperature, for metal dopants Hf, Ta, and Zr (in blue, orange, and green, respectively) under the same conditions as in Figure~\ref{fig-2}.}
     \label{fig-6}
\end{figure}

The defect concentrations of substitutional dopants (Hf\textsubscript{In}, Ta\textsubscript{In},  and Zr\textsubscript{In}) as a function of temperature for both \ce{In2O3} and \ce{(In_{0.5}Al_{0.5})_2O3} is illustrated in Fig.~\ref{fig-6}. Hf and Zr tend to have the largest concentrations, exceeding \( 10^{18} \text{ cm}^{-3}\) at 600\textdegree C in the case of In$_2$O$_3$ under oxygen-poor conditions. On the other hand, Ta exhibits much lower concentrations with a significant increase as temperature rises. Defect concentrations are also presented as a function of O chemical potential with an annealing temperature of [600-1000]\textdegree C, as shown in  Fig. S4.

The same trend as in the case group IV substitional defects is observed for the  transition metals considered in this work. Their concentrations increase as the oxygen chemical potential is decreased,  
All donor dopants in In$_2$O$_3$ exhibit significantly large concentrations. In all cases, alloying with Al leads to a reduction in defect concentration.

\begin{figure}[H]
     \centering
\includegraphics[width=10cm]{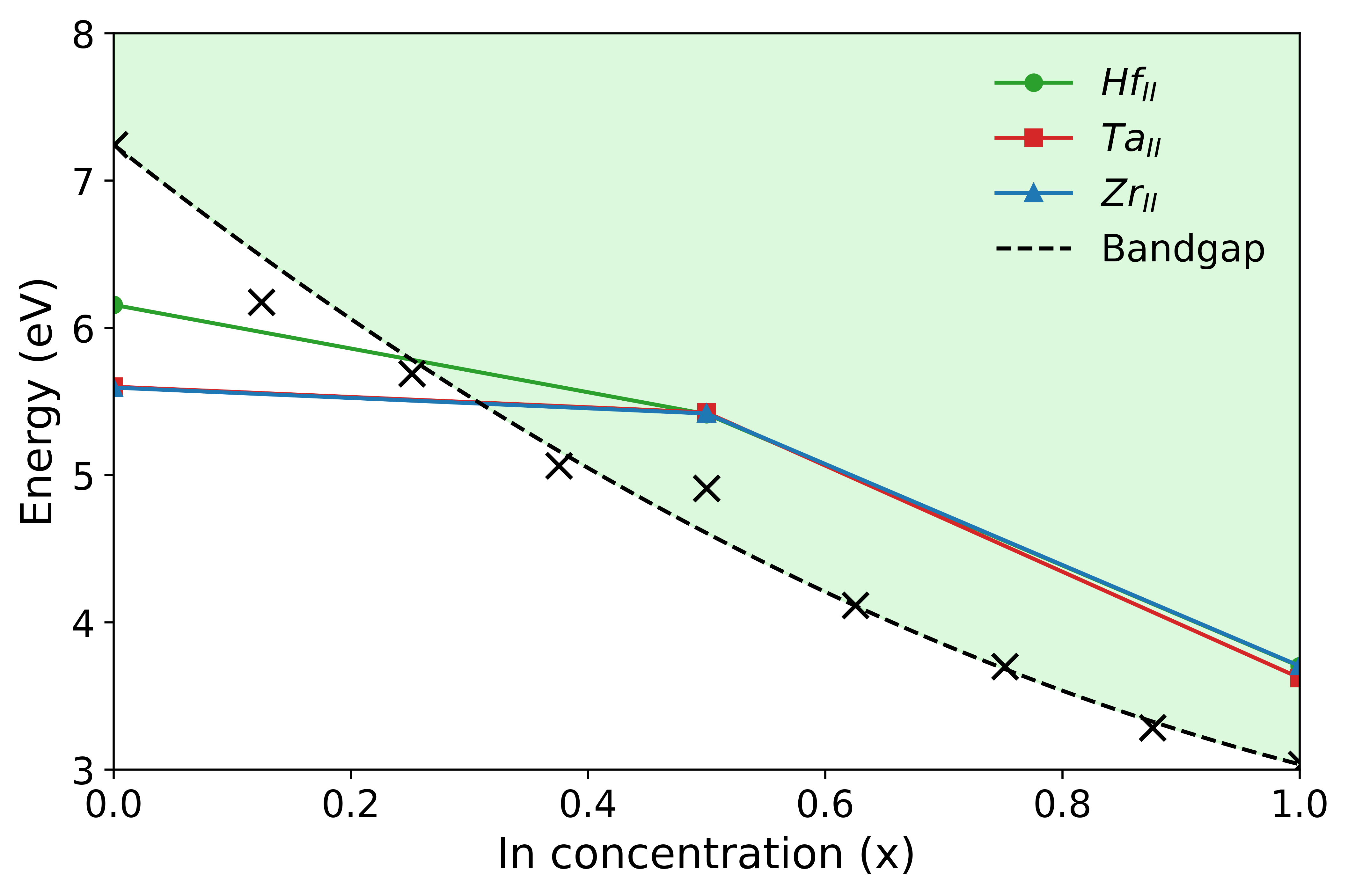} 
     \caption{First charge-state transition levels in \ce{(In_xAl_{1-x})_2O3} at$x=0$, $0.5$, and $1$, for metal dopants Hf, Ta, and Zr in octahedral sites (subscript II). Otherwise same as in Figure~\ref{fig-4}.}
     \label{fig-8}
\end{figure}

\begin{figure}[H]
     \centering
\includegraphics[width=16cm]{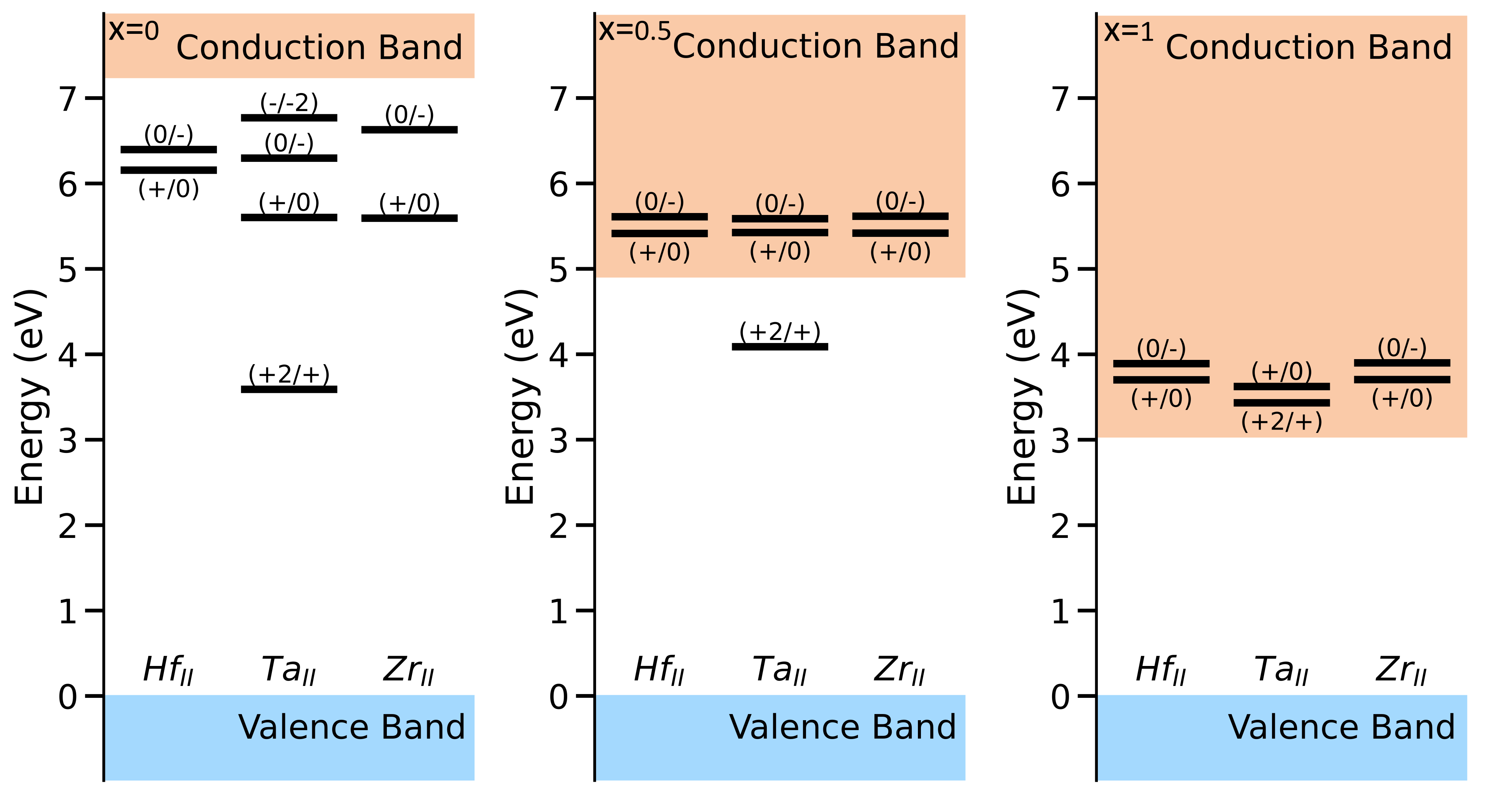} 
     \caption{The calculated charge-state transition levels for Hf, Ta, and Zr substitutional dopants in \ce{(In$_x$Al$_{1-x}$)$_2$O$_3$} alloys, for   $x=0$ (left), $x=0.5$ (center), and $x=1$ (right). Numerical values are provided in Supplementary Table~S4.}
     \label{figure-2}
\end{figure}

Fig.~\ref{fig-8} illustrates the energy at which the first charge-state transitions occur as a function of In content. 
In In$_2$O$_3$, all dopant transition levels occur above the CBM. 
When alloyed with Al, the Ta  dopant exhibits a deep donor (+2/+) charge-state transition within the band gap. 
In Al$_2$O$_3$, all dopants have charge-state transitions that occur within the band gap. 
For \( \mathrm{Hf}_{II} \), the critical composition occurs at \( x = 0.252 \), corresponding to a bandgap of 5.782~eV. \( \mathrm{Ta}_{II} \) and \(\mathrm{Zr}_{II} \)  on the other hand, exhibit almost similar critical compositions at \( x = 0.31 \) (bandgap = 5.5~eV). 
The charge-state transition levels for metal dopants are depicted in Fig.~\ref{figure-2}. 

\subsection{Acceptors (Mg, Zn, Cu)}

\begin{figure}[h]
     \centering
\includegraphics[width=17cm]{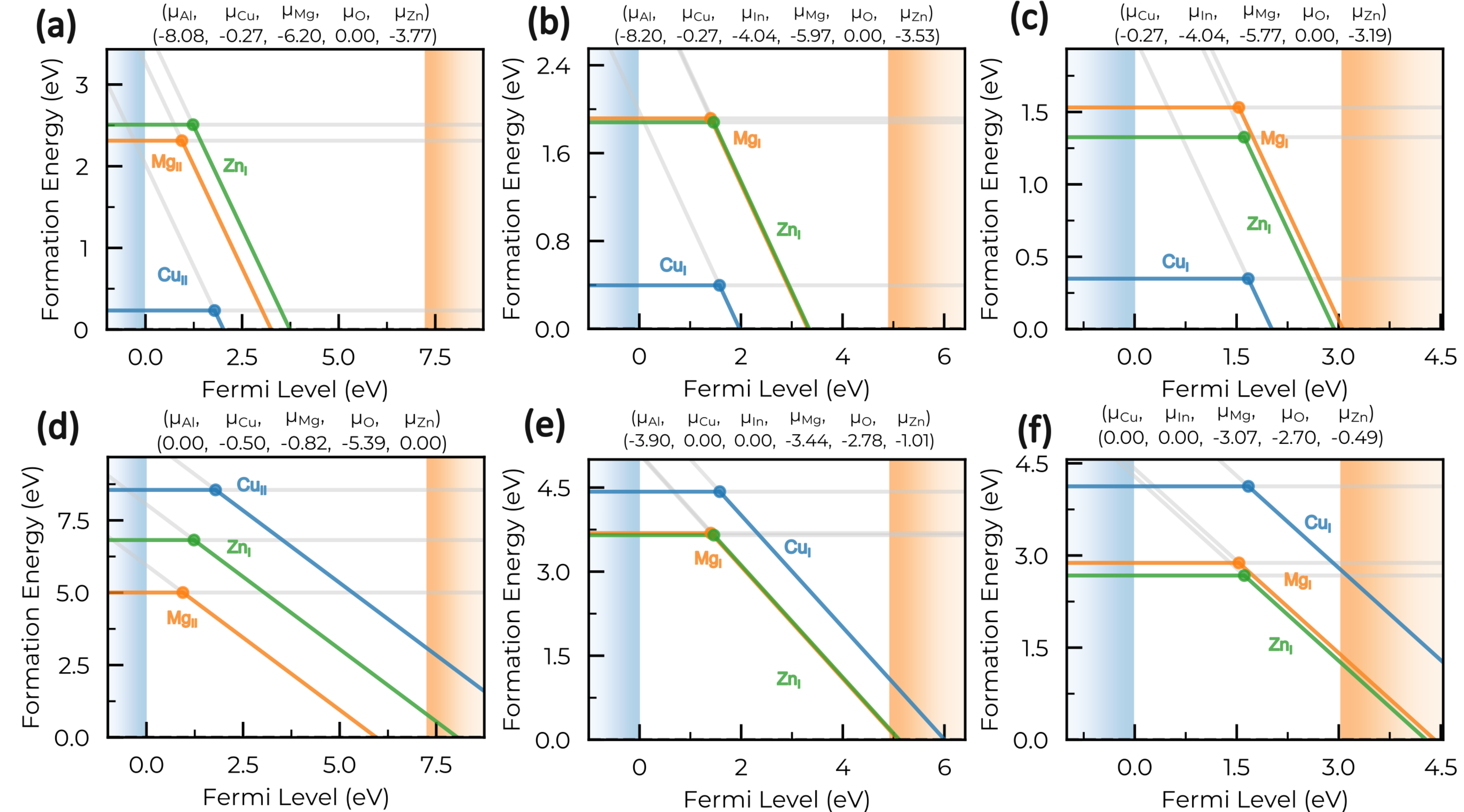} 
     \caption{As Figure~\ref{fig-1}, but for acceptor defects Cu, Mg, and Zn (in blue, orange, and green, respectively).}
     \label{fig-9}
\end{figure}

The calculated defect formation energies for Mg, Cu, and Zn acceptors are reported in Fig.~\ref{fig-9}  against the Fermi level. Only the most stable defect configurations were considered.

For both In$_2$O$_3$ and \ce{(In_{0.5}Al_{0.5})_2O3}, all substitutional acceptors preferentially occupy the tetrahedral site. However, under Al-rich conditions, the formation energies of both sites in the \ce{(In_{0.5}Al_{0.5})_2O3} ordered alloy are nearly identical. For Al$_2$O$_3$, Mg and Cu tend to occupy the octahedral site, instead. Under oxygen-rich conditions, Cu generally has the lowest formation energy, while Zn and Mg have almost similar formation energies, which remain slightly higher across the range of Fermi levels. Under oxygen-poor conditions, the formation energies of Zn and Mg dopants are slightly lower. 
Among the investigated dopants, Mg has the lowest formation energy  in Al$_2$O$_3$ under oxygen-poor conditions. The transition levels introduced by Cu, Mg, and Zn are all below the midgap, quite far from the VBM.  

\begin{figure}[h]
     \centering
\includegraphics[width=12cm]{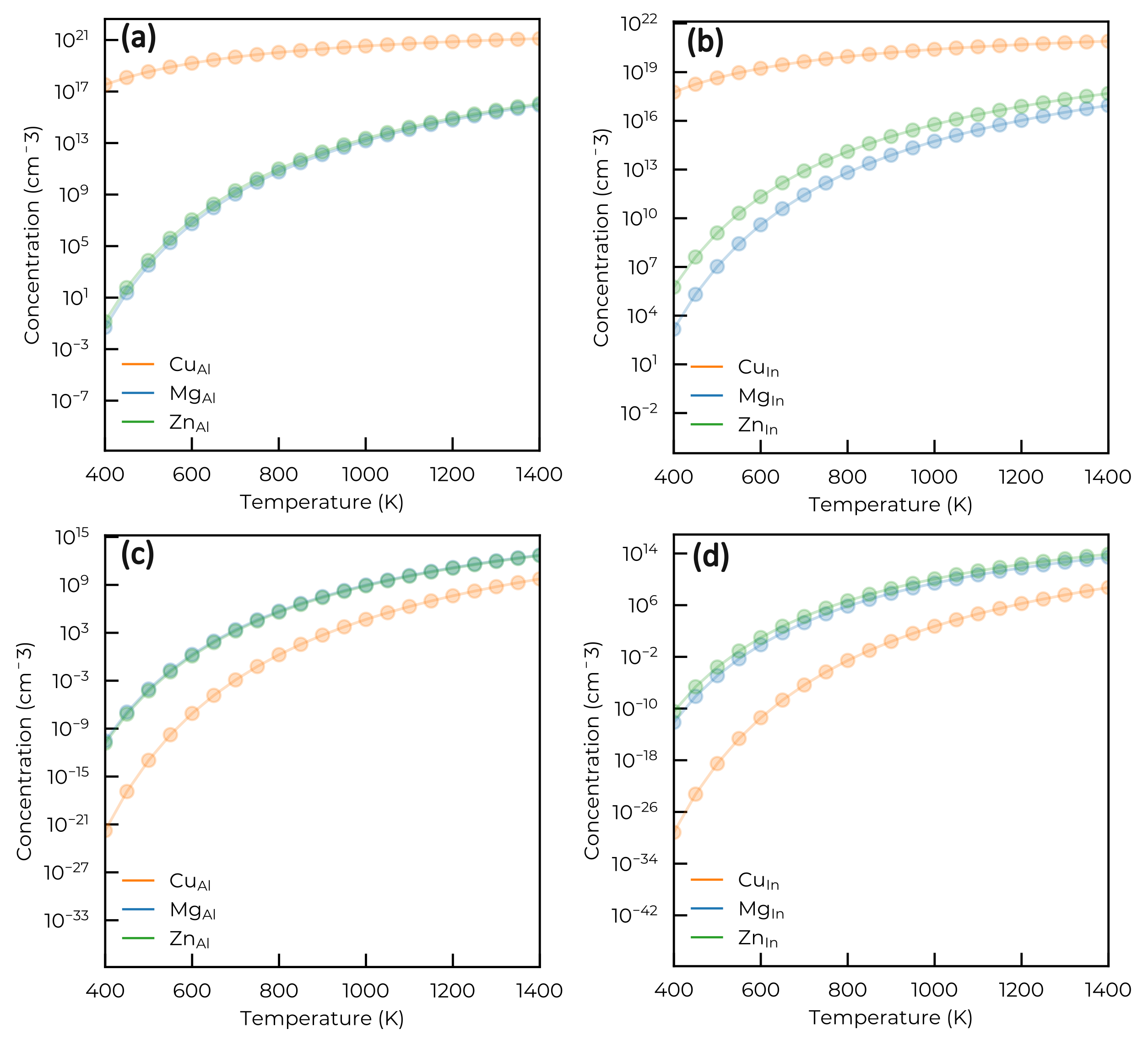} 
     \caption{Temperature sensitivity of the defect concentrations in \ce{(In_{0.5}Al_{0.5})_2O3} (a, c) and \ce{In2O3} (b, d) over an indicative range of temperature, for acceptor dopants Cu, Mg, and Zn (in blue, orange, and green, respectively) under the same conditions as in Figure~\ref{fig-2}.}
     \label{fig-10}
\end{figure}

\begin{figure}[H]
     \centering
\includegraphics[width=16cm]{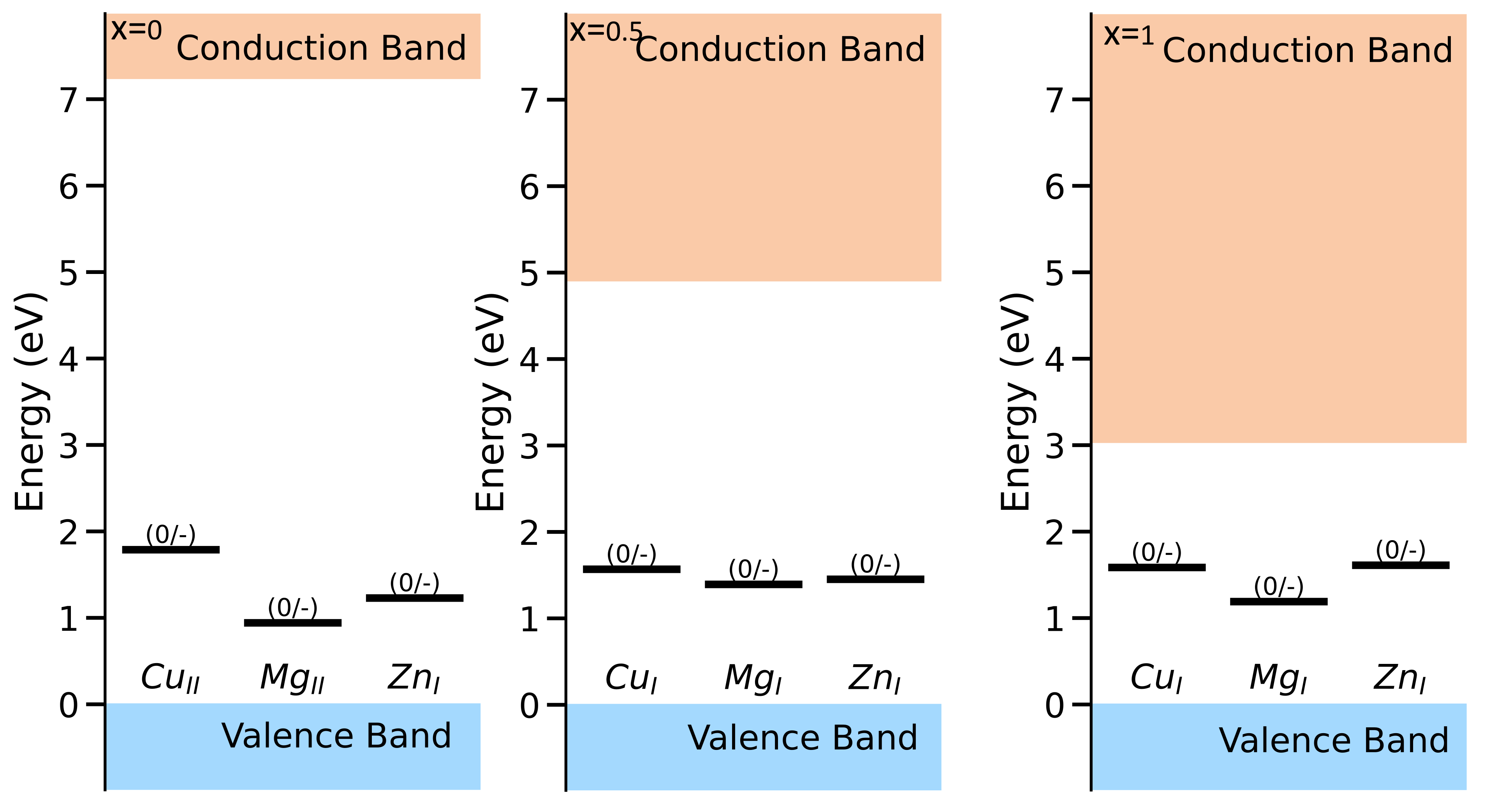} 
     \caption{The calculated charge-state transition levels for Cu, Mg, and Zn substitutional dopants in in \ce{(In$_x$Al$_{1-x}$)$_2$O$_3$} alloys, for   $x=0$ (left), $x=0.5$ (center), and $x=1$ (right). Numerical values are provided in Supplementary Table~S5.}
     \label{figure-3}
\end{figure}

The defect concentrations of substitutional dopants (Cu\textsubscript{In}, Mg\textsubscript{In},  and Zn\textsubscript{In}) as a function of temperature for both \ce{In2O3} and \ce{(In_{0.5}Al_{0.5})_2O3} is illustrated in Fig.~\ref{fig-10}.  
The defect concentrations in Al$_2$O$_3$ remain consistently low for all dopants (not presented), except for Cu dopants under extreme oxygen-rich conditions.  Consistent with the calculated formation energies shown in Fig.~\ref{fig-9}, for both In$_2$O$_3$ and \ce{(In_{0.5}Al_{0.5})_2O3} under oxygen-rich conditions, Cu substitution achieves the highest concentrations. Using the same annealing temperature ([600-1000]\textdegree C), the defect concentrations for acceptor dopants with variable O chemical potential are presented in  Fig.~S5. The charge-state transition levels for Cu, Mg, and Zn substitutional dopants are given in Fig.~\ref{figure-3}. All the investigated dopants create deep acceptor levels.

\newpage
\section{Discussion}

The monoclinic phase of $\mathrm{Al_2O_3}$, or $\theta$-$\mathrm{Al_2O_3}$, is highly anisotropic. Different coordination environments must be investigated,  as defects and dopants can be incorporated at multiple sites: two distinct sites for Al substitutions and three distinct sites for O substitutions.
Results presented in the previous Section are used to infer trends with composition and environmental conditions that can be observed experimentally. 
\subsection{Efficiency of Donors across Al composition and environmental conditions}

Group-IV elements, such as Si, Ge, Sn, and C, have been widely investigated as effective n-type dopants in UWBG materials.\cite{varley_oxygen_2010,lyons_carbon_2014,lany_defect_2018,bouzid_defect_2019,frodason_diffusion_2023} In agreement with previous investigations, all studied Group-IV elements preferentially occupy tetrahedral sites, except for Sn, which tends to occupy octahedral sites.\cite{varley_prospects_2020,seacat_achieving_2024}
Si is a well-known substitutional dopant, providing n-type conductivity in both GaN and Ga$_2$O$_3$.\cite{gordon_hybrid_2014,varley_prospects_2020}  However, Si substitution yields DX centres in UWBG materials like Al$_2$O$_3$, leading to a strong compensation effect that limits the efficiency of n-type doping at high Al compositions.\cite{lany_defect_2018,gordon_hybrid_2014,bouzid_defect_2019}
Sn and Ge donors have been reported to behave as donors when explicitly incorporated into Ga sites of Ga$_2$O$_3$, \cite{galazka_bulk_2014,lany_defect_2018} , with Sn being the favoured donor for commercial single crystals.  
The relatively high formation energies of C donors indicate a lower likelihood of incorporation when compared to other group IV elements.  C impurities have been predicted to unintentionally promote n-type conductivity in Ga$_2$O$_3$, as they are readily incorporated on Ga sites during Metal-Organic Chemical Vapour Deposition (MOCVD) growth, due to incomplete dissociation of metal-organic precursors.\cite{lyons_carbon_2014,mu_role_2022} 
In$_2$O$_3$ is well known for its n-type conductivity. \cite{bierwagen2010high,PhysRevMaterials.3.074604} Under oxygen-poor conditions, substitutional dopants like Sn$_{\text{In}}$ --- and, to a lesser extent, Si$_{\text{In}}$, Ge$_{\text{In}}$ --- lead to effective doping and larger n-type conductivity. Under oxygen-rich conditions, the formation energies of these dopants may still remain low enough to allow significant doping.

The defect concentrations shown in Fig.~\ref{fig-2} suggest that C dopants are not as effective in promoting free carriers in conduction band. While C dopant concentration increases with the rise of oxygen chemical potential, other defects exhibit an opposite trend. Overall, Si is a viable, yet less favorable, dopant compared to Ge and Sn. The trends observed in Fig.~\ref{fig-2} indicate that these dopants are more favorably incorporated under oxygen-poor conditions. 
In$_2$O$_3$ exhibits significantly higher dopant concentrations for all doping elements considered in this work. Alloying with Al yields a reduction in defect concentration, as shown in Fig.~\ref{fig-2} for the case of the ordered alloy, \ce{(In_{0.5}Al_{0.5})_2O3}. The defect concentration decreases dramatically, suggesting that a larger Al fraction in an (In$_x$Al$_{1-x}$)$_2$O$_3$ alloy is detrimental for n-type doping. 
For critical composition determination, we assumed a linear interpolation is reliable, in agreement with experimental observations for similar alloys, like AlGaN, and theoretical results for AlGaO alloys.\cite{mu_role_2022,gordon2014hybrid,skierbiszewski1999evidence,mccluskey1998metastability}  Moreover, calculated transition levels, as shown in Fig~\ref{figure-1}, and critical compositions, shown  in Fig~\ref{fig-4}, also indicate that higher Al composition limits the possibility of promoting n-type conductivity.


Transition metals including Hf, Ta and Zr have been already considered as alternative dopants in \ce{Ga2O3},\cite{cui_tuning_2019,saleh_electrical_2019,saleh_degenerate_2020, karbasizadeh_transition_2024} because of several imitations of group IV elements , like the formation of secondary phases at large Si concentrations, the high vapor pressure of Ge, or the elevated evaporation rates in case of Sn substitutions.\cite{villora_electrical_2008,galazka_bulk_2014,kuramata_high-quality_2016}
The low defect formation energies, shown in Fig.~\ref{fig-5}, especially at high In composition, suggest that Hf and Zr can indeed be considered as alternative dopants to group IV elements in \ce{(In_{x}Al_{1-x})_2O3}, as well. Ta dopant incorporation is comparably less favourable, given its relatively higher formation energies and deep transition level, as illustrated in  Figures \ref{fig-5} and \ref{figure-2}. In both \ce{In2O3} and \ce{(In_{0.5}Al_{0.5})_2O3}, under oxygen-poor conditions, substitutional dopants Hf$_{\text{In}}$ and Zr$_{\text{In}}$ display favourably low formation energies, facilitating their incorporation of these donors. For In$_2$O$_3$ under extreme In-rich conditions, these dopants maintain a low, nearly negative formation energy for all Fermi levels, suggesting high doping efficiency. The observed higher formation energy under oxygen-rich conditions remains low enough to allow non-negligible doping.

Defect concentrations, as presented in Fig.~\ref{fig-6}, follow the same trend as in the case group IV substitutional dopant. These concentrations increase as the oxygen chemical potential is decreased,  suggesting that Hf, Zr --- and, to lesser extend, Ta --- dopants are more favorably incorporated under oxygen-poor conditions.   Alloying with Al yields a reduction in defect concentration, as clearly shown in Fig. S4 for \ce{(In_{0.5}Al_{0.5})_2O3}) and In$_2$O$_3$. This finding suggests that a precise control of the Al composition will be required to achieve high doping concentration in practice, especially at higher Al composition. 

The suitability of metal dopants as alternatives to group IV elements is further assessed by determining the ionisation levels and the critical indium concentration, as shown in  Fig.~\ref{figure-2} and Fig.~\ref{fig-8}).
In the case of In$_2$O$_3$, transition levels occur above the CBM. It is then possible to enable n-type conductivity by explicitly incorporating Ta, Zr, and Hf dopants into In$_2$O$_3$. At higher Al compositions, dopant charge-state transitions occur within the band gap, limiting n-type conductivity promoted by these dopants. It is worth notating that, unlike group IV elements, transition metal dopants do not exhibit a direct (+/-) transition state, commonly associated with DX centers. 
When alloyed with Al, the Ta dopant exhibits a deep donor (+2/+) charge-state transition within the band gap. In Al$_2$O$_3$, all dopants have charge-state transitions that occur within the band gap, reflecting its persistent insulating behaviour.

In the case that a charge-state transition level --- \emph{e.g., }  the (+1/0) state for donors like Hf and Zr ---  lie in the conduction band, the defect is auto-ionized. It then acts as a shallow donor, spontaneously contributing a free electron to the conduction band without thermal activation. In this example, since there is no (+1/0) level within the band gap, large $n$-type doping and Fermi level pinning near the CBM will be obtained, if the defect formation energy is small. This situation can be beneficial in achieving low-resistivity n-type layers in power electronics devices like Schottky diodes or MOSFETs. However, wavefunctions of shallow donors are rather extended and a larger supercell may be required to confirm the nature of such states.

\subsection{Acceptor Dopant Potential for Compensation}

Exploring potential acceptor dopants is essential for addressing difficulties associated with p-type doping of \ce{Ga2O3} or \ce{(In_{x}Al_{1-x})_2O3} alloys. Introducing acceptor also enables the formation of more resistive layers, such as semi-insulating or interfacial layers in unintentionally n-doped materials. This could be achieved by incorporating suitable acceptor dopants to compensate the background donors. The most studied acceptor substitutional impurities for \ce{Ga2O3} include Mg, Cu and Zn. \cite{kananen_electron_2017,kyrtsos_feasibility_2018,gustafson_zn_2021,peelaers_deep_2019,yan_reducing_2021,cai_approach_2021,hommedal_broad_2024,luchechko_crystal_2024}
Under oxygen-poor conditions, the defect formation energies of these dopants reported in Fig.~\ref{fig-9} display a trend opposite to that observed for metals and group IV donors. In particular, the formation energies of the Mg, Cu and Zn defects are slightly larger. It is clear in both oxygen chemical potential conditions, all dopants remain effective in significantly reducing n-doping and preventing the Fermi level from shifting toward the CBM.

Based on the defect concentration shown in Fig.~\ref{fig-10}, under oxygen-rich conditions, dopant concentrations are generally higher compared to oxygen-poor conditions. This trend indicates that these dopants are more favourably incorporated under oxygen-rich conditions than the group IV and metal donors considered in the previous Section. Among acceptors, Cu concentration displays a steepest dependence on the O chemical potential, which is expected from its formation energy. Mg and Zn concentrations are comparably less sensitive to the O chemical potential. 
Along with the defect concentration, the location of the (0/-) transition level  also affects the effectiveness of a candidate acceptor. 
The charge-state transition levels for Cu, Mg, and Zn substitutional dopants presented in Fig.~\ref{figure-3} are consistent with what was previously found for other wide bandgap oxides \cite{cai_approach_2021}. Overall, these acceptors, especially under oxygen-rich conditions, can facilitate the creation of a semi-insulating and interfacial layer. On the other hand, achieving practical p-conductivity using these acceptors is ruled out. Our findings agree with the often reported difficulties in developing effective strategies to promote p-doping and hole mobilities in UWBG semiconducting oxides.\cite{McCluskey2020,peelaers_brillouin_2015}
Although in this study we have focussed on substitutional defects, other types of defects, including interstitial and intrinsic (\emph{e.g.} vacancies) defects, are possible in  \ce{(In_{x}Al_{1-x})_2O_3} alloys. To the best of our knowledge, they have not been investigated yet, and it would be interesting to compare their doping efficiency and compensating effect against those of the substitutional defects reported here.

\section{Conclusions}
We have performed atomistic modelling from first principles using the Heyd-Scuseria-Ernzerhof hybrid functional to investigate substitutional defects properties, including defect formation energies, defect densities, ionisation levels, and Al critical composition in \ce{(In$_x$Al$_{1-x}$)$_2$O$_3$} alloys. In particular, Si, Sn, C, and Ge (group IV), as well as promising alternative donors such as Ta, Zr, and Hf (transition metals) were investigated to determine their potential for n-type dopability of these alloys. Potential acceptors Mg, Zn, and Cu have also been considered. 

In our models, Hf and Zr show favourable properties, like low formation energy and shallow levels, as donors alternative to Si, especially under oxygen-poor conditions. These candidate donors can then be considered to overcome some of the group IV element limitations, including the formation of secondary phases at large Si concentrations, the high vapor pressure of Ge, or the elevated evaporation rates in case of Sn substitutions.
Our findings also suggest that acceptors Mg, Zn, and Cu --- while cannot promote useful p-doping in \ce{(In$_x$Al$_{1-x}$)$_2$O$_3$} alloys --- can be still beneficial for the compensation of unintentionally n-doped materials, 
\emph{e.g.}, in generating differently doped interfacial layers to improve rectification.\cite{splith_numerical_2021}
It is worth noting that the absence of an effective p-doping strategy in UWBG semiconducting sequioxides is a known technological issue. Along with the difficulties in finding shallow acceptors, large hole effective masses in the parent materials make efficient p-type conductivity unlikely to be achieved.\cite{McCluskey2020,peelaers_brillouin_2015}
As for \ce{Ga2O3}, resorting to heterojunctions could provide a more practical route to bipolar devices based on \ce{(In$_x$Al$_{1-x}$)$_2$O$_3$} alloys.
To this end, NiO is a material displaying favourable p-type characteristics, which has already been used in combination with \ce{Ga2O3} to demonstrate viable p-n junctions.\cite{pintor-monroy_tunable_2018,ezeh_wide_2023,lu_recent_2023,zhang_-ga2o3_2025}

\newpage


\begin{acknowledgements}
This work was supported by a research grant from the Department for the Economy Northern Ireland (DfE) under the US-Ireland R\&D Partnership Programme (USI 195).
Access to the computing facilities and support from the Northern Ireland High-Performance Computing (NI-HPC) service funded by EPSRC (EP/T022175), and to the UK national high-performance computing service, ARCHER2, through the UKCP consortium and funded by EPSRC (EP/X035891/1) are also gratefully acknowledged.
\end{acknowledgements}
\section*{Data Availability Statement}
Computational methodology and parameters are provided in the main manuscript and Supplemental Material~\cite{supplemental}. Additional data are available
from the corresponding author upon reasonable request.

\bibliography{reference}

\end{document}